\documentclass[journal]{IEEEtran}
\usepackage{graphicx}
\usepackage{graphicx,subfigure}
\usepackage{inputenc}
\usepackage[cmex10]{amsmath}
\usepackage{algorithmic}
\usepackage{algorithm}
\usepackage{amssymb}
\usepackage{mathrsfs}
\usepackage{stfloats}
\usepackage{dsfont}
\usepackage{setspace}
\usepackage{epstopdf}
\usepackage{epsfig}
\usepackage{balance}

\usepackage{enumitem}
\usepackage{pstricks}
\usepackage[noadjust]{cite}
\usepackage{filecontents}
\newtheorem{remark}{Remark}

\newtheorem{theorem}{Theorem}
\newtheorem{lemma}{Lemma}

\input epsf

\begin{document}
\title{Finite-SNR Analysis of Partial Relaying with Relay Selection in Channel-coded Cooperative Networks}
\author{Thang~X.~Vu,~\IEEEmembership{Member,~IEEE}, and Pierre~Duhamel,~\IEEEmembership{Fellow,~IEEE}
\thanks{T.~X.~Vu is with the Singapore University of Technology and Design, 487372 Singapore. E--mail: xuanthang\_vu@sutd.edu.sg.}
\thanks{P.~Duhamel is with Laboratory of Signals and Systems (LSS), French National Center for Scientific Research (CNRS) --  \'Ecole Sup\'erieure d'\'Electricit\'e (SUP\'ELEC) -- University of Paris--Sud 11, 91192 Gif--sur--Yvette, France. E--mail: pierre.duhamel@lss.supelec.fr.}}

\markboth{Submitted to IEEE Trans. Wireless Commun. May 2015}{Submitted paper}

\maketitle
\IEEEpeerreviewmaketitle

\begin{abstract}
This work studies the performance of a cooperative network which consists of two channel-coded sources, multiple relays, and one destination. Due to spectral efficiency constraint, we assume only one time slot is dedicated for relaying. Conventional network coding based cooperation (NCC) selects the best relay which uses network coding to serve two sources simultaneously. The performance in terms of bit error rate (BER) of NCC, however, is not available in the literature. In this paper, we first derive the closed-form expression for the BER of NCC and analytically show that NCC always achieves diversity of order two regardless the number of available relays and the channel code. Secondly, motivated by a loss in diversity in NCC, we propose a novel relaying scheme based on partial relaying cooperation (PARC) in which two best relays are selected, each forwarding half of the codeword to help one source. Closed-form expression for BER and system diversity order of the proposed scheme are derived. Analytical results show that the diversity order of PARC is a function of the operating signal-to-noise ratio (SNR) and the minimum distance of the channel code. More importantly, full diversity order in PARC can be achieved for practically operating finite SNRs with the proper channel code. Finally, intensive simulations present a huge SNR gain of PARC over NCC and reference schemes without relay selection.
\end{abstract}

\begin{keywords}
Cooperative diversity, relay selection, partial relaying, channel coding.
\end{keywords}
\section{Introduction}
In wireless networks, idle nodes have a potential to participate in transmission of other nodes to form cooperative communication. Cooperation among nodes has been shown as an effective technique to widen the coverage and to improve the performance of wireless networks in both terms of Signal-to-Noise Ratio (SNR) and diversity gain \cite{Sendonaris2003}. In the most basic cooperation form of single-source single-destination network, a relay estimates the source signal and then forwards it to the destination. It is shown that relay networks achieve a performance gain when compared with the non-cooperative counterpart \cite{Meulen1971}. In order to achieve this gain, however, additional orthogonal channel is usually required, which results in loses in spectral efficiency, especially when more than one relay is employed. Fortunately, such loss in multiple-relay networks can be effectively reduced by using opportunistic Relay Selection (RS), in which only the best relay is selected for cooperation \cite{Bletsas2006}. It is shown that RS can achieve full diversity order for single-source multiple-relay networks and outperform other relaying schemes in terms of SNR gain and effective capacity \cite{Lee2009}.

Network Coding (NC) has gained tremendous attention because of its potential improvement in diversity gain and throughput over classical routing techniques \cite{Ahlswede2000}. In NC, an intermediate  node combines multiple input packets into a linear combination which is then forwarded. Recently, there have been much studies on combining NC together with RS to further improve the spectral efficiency, mostly focusing on the Two-Way Relay Channel (TWRC). The authors in \cite{Yonghui2010} propose a joint design of NC with RS for Decode-and-Forward (DF) TWRC based on the max-min criterion in order to maximize the worst relay channel. In \cite{Jing2009}, a SNR-based suboptimal relay ordering is proposed for two way Amplify-and-Forward (AF) relay networks. A similar method is studied in \cite{Atapattu2013} to derive the system Outage Probability (OP), BER, and diversity order. Compared with research on RS in TWRC, which shows full diversity is achieved and is frequently available in the literature, research on RS in unidirectional relay networks is still limited. The study of NC with RS in unidirectional networks is first considered in \cite{Peng2008}. In this work the authors study the Diversity Multiplexing Tradeoff (DMT) and show that full diversity is achieved. However, the analysis in \cite{Peng2008} is relied on an unrealistic assumption that unintended packets are available at all destinations, which simplifies the unidirectional networks to TWRC. A generalized DMT analysis is presented in \cite{Topakkaya2011}. Likewise, the authors in \cite{Topakkaya2011} also assume an optimistic assumption that the selected channels are independent, whereas these channels belong to an ordered SNR sequence and hence are highly correlated \cite{Jing2009}. The analysis of the counterpart AF relaying in inter-user interference channels are studied in \cite{Ding2008,Geng2012,Guan2013}. We note that the diversity order in the above-mentioned works is studied via the limit of the upper bound of either OP or BER. Furthermore, most of these papers did not consider channel coding, which might be in contrast to practical scenarios in which nodes are usually protected by some forward error correction codes. 

In this paper, we investigate the performance of cooperative networks under practical conditions, i.e., the transmitted signals are protected by Convolutional Codes (CC). In the considered system, two sources communicate with a common destination with the aid of multiple available relays. This scenario can find applications in uplinks cellular mobile systems where two mobile users try to send data to the base station and some surrounding friendly, idle users can act as the relays. Due to the spectral efficiency constraint and processing delay limit, it is assumed that only one timeslot is dedicated for cooperation. The best RS is employed \cite{Bletsas2006} to effectively exploit the spatial diversity. For low-complexity functions at the relays, Demodulate-and-Forward (DMF) relaying protocol \cite{Cover1979} is used. At the destination, Cooperative Maximal Ratio Combining (C-MRC) detector \cite{Wang2007} is used priori to channel decoding to avoid error propagation. It is worth to note that C-MRC is a suboptimal detector and provides full diversity gain and a performance close to Maximum Likelihood (ML) receiver \cite{Nasri2013}. 

The contributions of the paper are as follows:
\begin{itemize}
	\item Firstly, we investigate the performance of Network Coding based Cooperation (NCC) in which one selected relay helps two sources by applying network coding on the estimated codewords. Closed-form expression of the BER is derived for all sources. From the analyzed BER, we analytically show that NCC always achieves a diversity of order two regardless the channel code and the total number of relays. This result coincides with the diversity order derived from OP analysis \cite{Topakkaya2011,Vu2015}.
	\item Secondly, we propose a novel relaying scheme named Partial Relaying based Cooperation (PARC). The key difference between PARC and NCC is that in the former two relays are selected for cooperation, each helping one source. Due to the spectral efficiency constraint, each selected relay in PARC forwards half of the estimated codeword to the destination.
	The cooperation based on partial relaying has been studied by some authors in \cite{Eckford2008,Khormuji2009}. Compared with these works, our proposed scheme has two main differences: i) we investigate the system via BER analysis, whereas these papers study the system through OP, which is fundamentally different from our method; and ii) the proposed PARC employs RS to improve the spectral efficiency, while these papers do not.
	\item Thirdly, insightful theoretical analysis is provided for PARC in finite SNR regime. Particularly, closed-form expression for BER and the diversity order are derived, which reveals that the \emph{instantaneous} diversity order of PARC is a function of the operating SNR and the minimum distance of the channel code\footnote{To investigate the system in finite SNR regime, instantaneous diversity order is defined as the generalized definition of classical diversity order at any SNR value. More details are presented in \ref{sec:PARC_Diversity}}. More importantly, PARC can achieve full diversity order with suitable channel codes in low and mediate SNR regime, the operating SNR region in practical systems. Intensive simulation results show a large SNR gain of PARC over NCC and other reference schemes.
\end{itemize}

The rest of the paper is organized as follows. Section~\ref{sec:SystemModel} describes the system model of PARC and NCC. Section~\ref{sec:Selection} provides details for the relay selection process. Section~\ref{sec:PerformancePARC} analyzes BER and diversity order of PARC. The performance analysis of NCC is analyzed in Section \ref{sec:PerformanceNCC}. Section~\ref{sec:Results} shows numerical results. Finally, conclusions and discussions are given in Section~\ref{sec:Conclusions}.
\section{System Model} \label{sec:SystemModel}

The system under consideration consists of two sources denoted by $\rm S_1$ and $\rm S_2$, $N_r$ relays denoted by $\mathrm{R}_i$ with $1 \leq i \leq N_r$, and one destination denoted by D. All nodes are equipped with a half-duplex single antenna. The system is assumed to operate on orthogonal channels with perfect time synchronization. As a result, a cooperation period is divided into two phases: broadcast phase and relaying phase. Due to the spectral efficiency constraint and processing time limit, we assume that only one time slot is dedicated for relaying. All the channels are subject to block Rayleigh fading plus Additive White Gaussian Noise (AWGN). In order to minimize the computation at the relays, Demodulate-and-Forward (DMF) relaying protocol is used. To effectively exploit spatial diversity and to achieve high spectral efficiency, relay selection is used \cite{Bletsas2006}. The relay selection process is performed at the beginning of every cooperation period and will be described in details in Section~\ref{sec:Selection}. 
\subsection{Partial Relaying based Cooperation (PARC)}

\begin{figure*}[!t]
	\normalsize
	\hspace{0.3cm}
	\subfigure[]{\includegraphics[width = 0.45\textwidth]{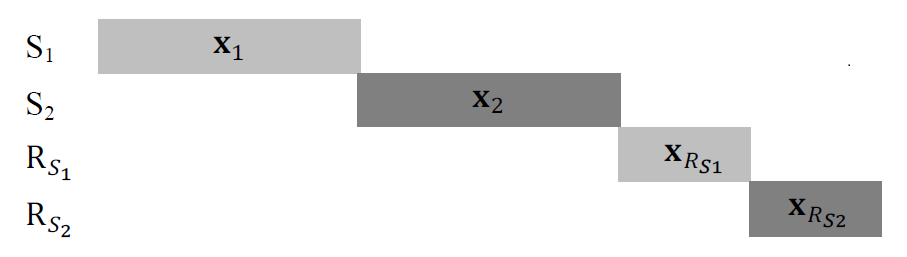}}
	\hspace{0.8cm}
	\subfigure[]{\includegraphics[width = 0.45\textwidth]{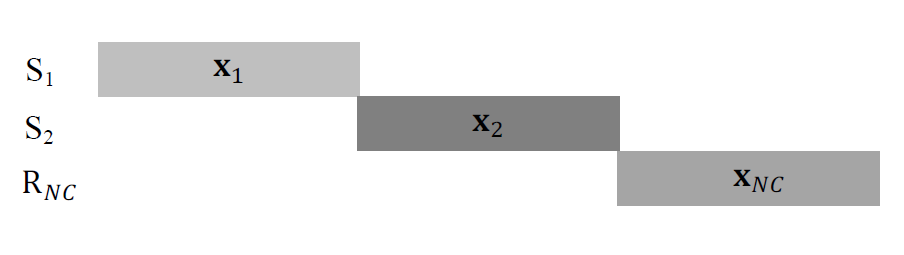}}
	\caption{Time allocation for the Partial Relaying based Cooperation (a) and Network Coding based Cooperation (b). In PARC, two relays are selected out of $N_r$ total number of available relays, each forwarding half of source codeword. On the other hand, only one relay is selected in NCC. The selected relay $\mathrm{R}_{NC}$ in NCC forwards the whole network-coded codeword to help two sources simultaneously.}
	\label{fig:SystemModel}
\end{figure*}

Motivated by our previous work which showed full diversity gain is achieved for three-node relay networks in low and medium SNRs even when the relay only forward parts of the codeword \cite{Vu2013}, we propose to combine relay selection and partial relaying. In the proposed PARC, two relays are selected in which each is the best relay for one source. Since two relays are active in the relaying phase, each relay only occupies half of relaying time slot, as shown in Fig.~\ref{fig:SystemModel}a. Consequently, the selected relay can only forward half of the estimated codeword to the destination. Although forwarding half of the estimated codeword, full diversity gain in finite SNRs\footnote{It is also called instantaneous diversity order, which is measured as the slope of the BER curve plotted in log-log scale as a function of SNR.} is expected when suitable channel code is employed, as shown later on in Section~\ref{sec:PerformancePARC}. 

First, source $\mathrm{S}_i, i = 1,2$, encodes a $K$-length data message $\mathbf{u}_i$ into a codeword $\mathbf{c}_i$ which contains $N$ coded symbols by a convolutional code $\mathbf{g}$ with code rate $K/N$. The codeword $\mathbf{c}_i$ is then modulated into a signal $\mathbf{x}_i$. Next, the signal $\mathbf{x}_i$ is broadcasted to the relays and the destination. Denote $\mathrm{R}_{S_1}, \mathrm{R}_{S_2}$ as the selected relays for $\rm S_1$ and $\rm S_2$, respectively. The received signal at the destination and the selected relays at the end of first phase are given as follows:
\begin{align} \label{eq:1}
\left\{ \begin{array}{ll}
  \mathbf{y}_{S_iR_{S_i}} &= \sqrt{P_{S_iR_{S_i}}} h_{S_iR_{S_i}} \mathbf{x}_i + \mathbf{n}_{S_iR_{S_i}}, i = 1,2, \\
  \mathbf{y}_{S_iD} &= \sqrt{P_{S_iD}} h_{S_iD} \mathbf{x}_i + \mathbf{n}_{S_iD}, i = 1,2,
  \end{array}
  \right.
\end{align}
where $P_{XY}$ with $X \in \{S_1, S_2\}, Y \in \{R_{S_1}, R_{S_2}, D\}$ is the average received power at node $Y$ from node $X$, including the path loss; $h_{XY}$ is the channel fading coefficient between $X$ and $Y$, which is a complex Gaussian random variable with zero mean and unit variance, i.e., $\mathbb{E}\left\{|h_{XY}|^2\right\}=1$, and is mutually independent among $X \to Y$ channels; $\mathbf{n}_{(.)}$ is a noise vector whose components are Gaussian random variables with mean zero and variance $\sigma^2$.

At the end of the first phase, the selected relay estimates the source coded symbols and forwards them to the destination. In the proposed scheme, the selected relay $\mathrm{R}_{S_i}, i = 1,2,$ uses half of the relaying time slot to forward half of the codeword $\mathbf{c}_i$ to the destination. More specifically, the selected relay $\mathrm{R}_{S_i}$ first estimates $L = \lfloor N/2 \rfloor$ source coded symbols to form an estimated punctured codeword $\mathbf{\hat{c}}_{R_{S_i}} = \{\hat{c}_{R_{S_i},l}\}_{l\in \Theta}$, where $\lfloor a \rfloor$ denotes the largest integer less than $a$, and $\Theta = \{k_1,\ k_2,\dots, k_L\}$ being the set of indexes of the symbols which are helped by the relay $\mathrm{R}_{S_i}$. The index set $\Theta$ are determined randomly\footnote{Other selection of $\Theta$, e.g., optimal index set, can be employed, but are beyond the scope of this paper.}. The source coded symbols at the relay are estimated by the ML detector as follows:
\begin{align}
\hat{c}_{R_{S_i},l} = \arg\min_{c_{i,k_l}\in\{0,1\}} \{|y_{S_iR_{S_i},k_l} - \sqrt{P_{S_iR_{S_i}}}h_{S_iR_{S_i}} x_{i,k_l}|^2\}, \notag 
\end{align}
$\forall k_l \in \Theta$, where $x_{i,k_l}$ being the corresponding modulated symbol of $c_{i,k_l}$. Next, $\mathrm{R}_{S_i}$ modulates $\hat{\mathbf{c}}_{R_{S_i}}$ into the modulated signal $\hat{\mathbf{x}}_{R_{S_i}}$ and then forwards it along with the index set $\Theta$ to the destination. The cost for conveying the index set is negligible since it can send, e.g., the seed of the random interleaver, to the destination. 

The received signal at the destination transmitted from the relay is given as:
\begin{eqnarray}\label{eq:2}
\mathbf{y}_{R_{S_i}D}=\sqrt{P_{R_{S_i}D}}h_{R_{S_i}D}\hat{\mathbf{x}}_{R_{S_i}} + \mathbf{n}_{R_{S_i}D}, i = 1,2,
\end{eqnarray}
where $h_{R_{S_i}D}$ is the channel coefficient from $\mathrm{R}_{S_i} \to \mathrm{D}$, and $\mathbf{n}_{R_{S_i}D}$ is a noise vector whose components are Gaussian random variable with zero mean and variance $\sigma^2$.

After receiving two signals from the source and the relay, the destination starts the decoding process with two consecutive steps: demodulating and decoding. Assuming that the CSI of all channels, i.e., $\mathrm{S}_i \to \mathrm{D}, \mathrm{S}_i \to \mathrm{R}_{S_i}$ and $\mathrm{R}_{S_i} \to \mathrm{D}$ channels, are available at the destination, the destination first applies the C-MRC detector \cite{Wang2007} to demodulate the coded bits for source $\mathrm{S}_i, i = 1,2$, as follows:
\begin{align}
\hat{c}_{i,k} = \arg\min_{c_{i,k}\in\{0,1\}} \mathcal{M} (x_{i,k}),\ 1 \leq k\leq N, \notag
\end{align}
where the detection metric $\mathcal{M} (x_{i,k}) = |y_{S_iD,k} - \sqrt{P_{S_iD}} h_{S_iD}x_{i,k} |^2$ if $k\, \notin \,\Theta$; otherwise 
\begin{align}\label{eq:3}
\mathcal{M}(x_{i,k}) ~=&~  \left|y_{S_iD,k} - \sqrt{P_{S_iD}} h_{S_iD} x_{i,k}\right|^2 \\
&~+ \lambda_{R_{S_i}} \left| y_{R_{S_i}D,k} - \sqrt{P_{R_{S_i}D}} h_{R_{S_i}D} \hat{x}_{R_{S_i},k}\right|^2.\notag
\end{align}
In \eqref{eq:3}, $\lambda_{R_{S_i}}$ is the parameter of the C-MRC detector which is computed as $\lambda_{R_{S_i}} \triangleq  \frac{\min (\gamma_{S_iR_{S_i}}, \gamma_{R_{S_i}D})}{\gamma_{R_{S_i}D}}$, where $\gamma_{XY} =  P_{XY} |h_{XY}|^2/\sigma^2$ being the instantaneous SNR of the channel $X \rightarrow Y$. 

The C-MRC detector then computes log-likelihood ratio values of the coded bits and sends them to the channel decoder. Finally, the channel decoder applies the BCJR algorithm \cite{Bahl1974} to decode the transmitted data. 

\begin{remark}
	In our protocol, the selected relay always forwards the estimated symbols to the destination. Fortunately, possible decoding error in $\hat{c}_{R_{S_i},l}$, hence error propagation, is effectively mitigated by $\lambda_{R_{S_i}}$ in C-MRC. For example, if the source-relay channel is too noisy, i.e., $\gamma_{S_iR_{S_i}}$ is too small, it is highly probable that $\mathrm{R}_{S_i}$ decodes with errors. In this case, however, $\lambda_{S_i}$ is small and the contribution of the relayed signal is negligible.  
\end{remark}
\subsection{Network Coding based Cooperation (NCC)}\label{sec:NCC}

In NCC, the relays use network coding to help both sources simultaneously to improve the spectral efficiency. One cooperation in NCC is also divided into two phases: broadcast phase and relaying phase. The broadcast phase is similar to that in PARC, whereas in the relaying phase, only one best relay is active. Time allocation of NCC is depicted in Fig.~\ref{fig:SystemModel}b. Unlike PARC, the selected relay in NCC forwards the whole network-coded codeword to the destination. Without loss of generality, denote by $\mathrm{R}_{NC}$ the selected relay in NCC. The received signal at $\mathrm{R}_{NC}$ is given as follows:
\begin{align} \label{eq:NC1}
\mathbf{y}_{S_iR_{NC}} = \sqrt{P_{S_iR_{NC}}} h_{S_iR_{NC}} \mathbf{x}_i + \mathbf{n}_{S_iR_{NC}}, i = 1,2. 
\end{align}
At the end of the first phase, $\mathrm{R}_{NC}$ decodes the estimate $\hat{\mathbf{x}}_{iR}$ of $\mathbf{x}_i, i = 1,2$, using the ML detector as follows:
\begin{align}
\hat{c}_{iR,k} = \arg\min_{c_{i,k}\in\{0,1\}} \{|y_{S_iR_{NC},k} - \sqrt{P_{S_iR_{NC}}}h_{S_iR_{NC}} x_{i,k}|^2\}&,\notag \\
i \in \{1,2\},~ 1 \leq k  \leq N&, \notag
\end{align}
where $x_{i,k}$ being the corresponding modulated symbol of $c_{i,k}$. Then $\mathrm{R}_{NC}$ performs network encoding to get $\hat{\mathbf{c}}_{NC}$, where $\hat{c}_{NC,k} = \hat{c}_{1R,k} \oplus \hat{c}_{2R,k},\ \forall k$, and $\oplus$ denotes the binary XOR operation.

The received signal at the destination from the selected relay is given by:
\begin{align}\label{eq:11}
\mathbf{y}_{R_{NC}D} = \sqrt{P_{R_{NC}D}}h_{R_{NC}D}\hat{\mathbf{x}}_{NC} + \mathbf{n}_{R_{NC}D},
\end{align}
where $\hat{\mathbf{x}}_{NC}$ is the modulated signal of $\hat{\mathbf{c}}_{NC}$. After two phases, the destination receives three channel observations from two sources and the selected relay. To decode the source data, the destination applies joint network/channel decoding algorithm to a "compound code" $\mathbf{G}$ \cite{Vu2013} which sees the relayed signal as additional parity bits (redundancy). The compound code $\mathbf{G}$ is formed from the individual code $\mathbf{g}$ as follows:
\begin{align} \label{eq:G}
\mathbf{G} = \left[\begin{array}{ccc}
\mathbf{g} & \mathbf{0} & \mathbf{g}\\
\mathbf{0} & \mathbf{g} & \mathbf{g}
\end{array} \right],
\end{align}
where $\mathbf{0}$ is a zero matrix withs same size as $\mathbf{g}$. For full details of joint decoding at the destination, we refer the readers to \cite{Vu2013}.
\section{Relay Selection for PARC and NCC}\label{sec:Selection}
In this section, we describe in details the relay selection processes for PARC and NCC and provide essential characteristics of the selected relay channels. 

\subsection{Relay Selection in PARC} \label{sec:RelaySelectionPARC}
The relay selection is based on the suboptimal max-min criterion that maximizes the worst end-to-end SNR and reduces computational complexity \cite{Yonghui2010}. The relay selection process in PARC is executed for each source separately and can be done in distributed manner similar to \cite{Bletsas2006}. After the channel estimation, the relays set a timer that is inversely proportional to their channel gain. The first relay whose timer is zero will send a pulse to the destination. Upon receiving the pulse, the destination declares the chosen relay. Because the selection procedure for two sources are similar, we avoid source subscript in this subsection for notation brevity. Particularly, the source is denoted by $\rm S$ and the selected relay is denoted by $\mathrm{R}_S$. Since the relayed symbols received at the destination using DMF protocol can be well described by \textit{equivalent channel} \cite{Bletsas2006}, we model a two-hop source-relay-destination link by an equivalent single-hop channel $\gamma_j = \min \left\{\gamma_{SR_j},\gamma_{R_jD}\right\}$, $1 \leq j \leq N_r$. In Rayleigh fading channel, both $\gamma_{SR_j}$ and $\gamma_{R_jD}$ are exponential random variables with mean $\overline{\gamma}_{SR_j}$ and $\overline{\gamma}_{R_jD}$, respectively. Using the property of the Min function, it is straightforward to show that $\gamma_j$ is also an exponent random variable with mean $\overline{\gamma}_j$, which is computed as:
\begin{align*}
	\frac{1}{\overline{\gamma}_j} = \frac{1}{\overline \gamma_{SR_j}}+ \frac{1}{\overline \gamma_{R_jD}}.
\end{align*}
To minimize possible errors of the relayed symbols, the relay that has the biggest equivalent channel is selected for cooperation:
\[
\mathrm{R}_S = \arg \max_{R_j} \gamma_j,~ 1 \leq j \leq N_r.
\]
The equivalent channel of the selected relay, $\gamma_{Sel}$, is given by:
\begin{align*}
	\gamma _{Sel} = \max \{ \gamma_1, \dots ,\gamma_{N_r} \}.
\end{align*}
By using the Max function \cite{Simon2005}, the Probability Density Function (PDF) of $\gamma_{Sel}$ is given in a shorten form as follows:
\begin{align}
f_{\gamma_{Sel}}\left(\gamma \right)\! =\!  {\mathop \sum \limits_{j=1}^{N_r}} \Big( (-1)^{j-1}\! \mathop \sum \limits_{\begin{subarray}{c}
	{n_1}=1,\dots,{n_j} = 1 \\
	{n_1} \neq \dots \neq {n_j}
\end{subarray}} ^{N_r} \frac{1}{\overline{\gamma}_{Sel,j}} \exp\Big(\!-\frac{\gamma}{\overline{\gamma}_{Sel,j}}\Big)\Big), \notag
\end{align}
where
\begin{align*}
\frac{1}{\overline{\gamma}_{Sel,j}} = \sum \limits_{k=n_1}^{n_j} \left( \frac{1}{\overline \gamma_{SR_{k}}} + \frac{1}{\overline \gamma_{R_{k}D}}\right).
\end{align*}

The moment generating function (MGF) of $\gamma_{Sel}$ is given by:
\begin{align}\label{eq:4}
\Psi_{\gamma_{Sel}}(s) = {\mathop \sum \limits_{j=1}^{N_r}} \Big((-1)^{j-1} \mathop \sum \limits_{\begin{subarray}{c}
		{n_1}=1,\dots,{n_j} = 1 \\
		{n_1} \neq \dots \neq {n_j}
	\end{subarray}} ^{N_r} \frac{1}{1 + \overline{\gamma}_{Sel,j}s}\Big).
\end{align}
\begin{remark}
	In the proposed PARC, the relay selection is performed for each source separately. Also the decoding at the destination is executed separately for each source. 
\end{remark}
\subsection{Relay Selection in NCC}
\label{sec:RelaySelectionNCC}
The relay selection process is performed in NCC based on a criterion that minimize possible error of network-coded symbols. Because an error of the network-coded signal can result from either source-relay links or relay-destination link, the network-coded symbols can be seen as if it has been transmitted via an equivalent channel which yields the same error probability \cite{Nasri2013}. Using the equivalent error probability for network-coded symbols, the two-hop source-relay-destination channel corresponding to the relay $\mathrm{R}_j$ can be tightly modeled as follows \cite{Nasri2013}:
\begin{align*}
\gamma_{eq,j} = \min \{\gamma_{S_1R_j},\gamma_{S_2R_j}, \gamma_{R_jD}\}.
\end{align*}
Because $\gamma_{S_1R_j},\gamma_{S_2R_j}$, and $\gamma_{R_jD}$ are exponential random variables, it is straightforward to show that $\gamma_{eq,j}$ is also an exponential random variable whose mean $\overline{\gamma}_{eq,j}$ is given by
\begin{align*}
\frac{1}{{\overline \gamma }_{eq,j}} = \frac{1}{{\overline \gamma }_{S_1R_j}} + \frac{1}{{\overline \gamma }_{S_2R_j}} + \frac{1}{{\overline \gamma }_{R_jD}}.
\end{align*}
The best relay in NCC, denoted by $\mathrm{R}_{NC}$, is selected by the max-min criterion as:
\begin{align*}
\mathrm{R}_{NC} = \arg \max_{R_j} \gamma_{eq,j},~ 1 \leq j \leq N_r.
\end{align*}
The equivalent network-coded channel of the selected relay is chosen as follows:
\begin{align*}
{\gamma _{NC}} = \max \{ {\gamma _{eq,1}}, \dots ,{\gamma _{eq,N_r}}\}.
\end{align*}
Because the $\gamma_{eq,j}$ are mutually independent, the Cumulative Density Function (CDF) of $\gamma_{NC}$ is computed as: $F_{\gamma_{NC}}(\gamma) = \prod_{j=1}^{N_r} F_{\gamma_{eq,j}}(\gamma)$. Taking the derivative of $F_{\gamma_{NC}}(\gamma)$ we obtain the PDF of $\gamma_{NC}$ expressed in the simplified form as follows:
\begin{align*} 
f_{\gamma_{NC}}&\left(\gamma \right) = \\
&{\mathop \sum \limits_{j=1}^{N_r}} \Big( (-1)^{j-1} \mathop \sum \limits_{\begin{subarray}{c}
	{n_1}=1,\dots,{n_j} = 1 \\
	{n_1} \neq \dots \neq {n_j}
	\end{subarray}} ^{N_r} \frac{1}{\overline{\gamma}_{NC,j}} \exp\Big(-\frac{\gamma}{\overline{\gamma}_{NC,j}}\Big)\Big),\notag
\end{align*}
where
\begin{align*}
\frac{1}{\overline{\gamma}_{NC,j}} =\sum \limits_{k=n_1}^{n_j} \left( \frac{1}{\overline \gamma_{S_1R_k}} + \frac{1}{\overline \gamma_{S_2R_k}} + \frac{1}{\overline \gamma_{R_kD}} \right).
\end{align*}
The MGF of $\gamma_{NC}$ is calculated as follows:
\begin{align} \label{eq:12}
\Psi_{\gamma_{NC}}(s) = {\mathop \sum \limits_{j=1}^{N_r}} \Big( (-1)^{j-1} \mathop \sum \limits_{\begin{subarray}{c}
	{n_1}=1,\dots,{n_j} = 1 \\
	{n_1} \neq \dots \neq {n_j}
	\end{subarray}} ^{N_r} \frac{1}{1 + \overline{\gamma}_{NC,j} s} \Big).
\end{align}

\section{Performance Analysis for Partial Relaying based Cooperation} \label{sec:PerformancePARC}
In this section, we analyze BER and diversity order of PARC using the equivalent channel model. Because the decoding procedure for two sources is similar, the analysis for two sources is analogy. To simplify notation in the analysis, we drop source subscript. After two phases, the destination receives two signal from $\rm S$ and $\mathrm{R}_S$, the selected relay. The combined signal at the output of C-MRC detector can be classified into two groups: the first group consists of symbols which are helped by the selected relay, and the second group includes the rest symbols which are not relayed. In other words, the received signal at the destination can be seen as an output of block fading channel with 2 blocks: one block consisting of $N-L$ symbols only sees the channel $\gamma_{SD}$, and the other one which contains $L$ symbols sees both channel $\gamma_{SD}$ and channel $\gamma_{Sel}$. 
\subsection{Bit Error Rate Analysis}
Let $\mathrm{P_u}(d)$ be the Unconditioned Pair-wise Error Probability (UPEP)\footnote{The unconditioned pair-wise error probability does not depend on the fading channels.}, which is probability that the destination decodes for a codeword with the Hamming distance $d$ (number of non-zero coded bits in $\mathbf{c}_i$) when the all-zero codeword was transmitted. The BER of PARC is upper-bounded as follows \cite{Glavieux2007}:
\begin{eqnarray}\label{eq:5}
	\mathrm{Pe} \leq \sum_{d=f}^{N} w(d)\mathrm{P_u}(d),
\end{eqnarray}
where $f$ is the minimum distance of the channel code $\mathbf{g}$, and $w(d)$ is input weights which is number of non-zero information bits in $\mathbf{u}_i$ and is computed directly from structure of the code. The UPEP $\mathrm{P_u}(d)$ is the expectation over the channel fading coefficients of the Conditioned Pair-wise Error Probability (CPEP) $\mathrm{P_c}(d)$: $\mathrm{P_u}(d) = \mathbb{E}\{\mathrm{P_c}(d)\}$. The CPEP $\mathrm{P_c}(d)$ obviously depends on the fading channels and how $d$ non-zero coded bits are distributed on the two blocks ($\gamma_{SD}$ and $\gamma_{SD}+\gamma_{Sel}$). Denote $\mathbf{D}_d = \{d_1, d_2\}$, $d_1 +d_2 = d$, as the weight pattern that presents how $d$ weights are distributed on the two blocks. Because $d$ non-zero coded bits uniformly locate in the two blocks, the CPEP can be further analyzed as follows:
\begin{align}\label{eq:6}
	\mathrm{P_c}(d) = \sum_{\mathbf{D}_d} \mathrm{P_c}(d|\mathbf{D}_d) p(\mathbf{D}_d),
\end{align}
where $p(\mathbf{D}_d)$ is the probability of pattern $\mathbf{D}_d$, which is computed by combinatoric computation as
\begin{align*}
	p(\mathbf{D}_d) = \frac{\mathcal{C}^{N-L}_{d_1} \times \mathcal{C}_{d_2}^{L}}{\mathcal{C}^N_d},
\end{align*}
where $\mathcal{C}^n_k = \frac{n!}{(n-k)!\times k!}$.

Substituting \eqref{eq:6} into $\mathrm{P_u}(d)$ we obtain:
\begin{eqnarray}\label{eq:7}
	\mathrm{P_u}(d) = \sum_{\mathbf{D}_d} \underbrace{\mathbb{E}\left\{\mathrm{P_c}(d|\mathbf{D}_d)\right\}}_{\mathrm{P_u}(d|\mathbf{D}_d)} p(\mathbf{D}_d).
\end{eqnarray}
Given the pattern $\mathbf{D}_d = \{d_1, d_2\}$, there are $d_1$ non-zero coded bits undergoing through block $\gamma_{SD}$ and $d_2$ non-zero coded bits undergoing through block $\gamma_{SD} + \gamma_{Sel}$. As a result, the CPEP $\mathrm{P_c}(d|\mathbf{D}_d)$ is calculated, by using the same techniques in \cite{Vu2013a}, as follows:
\begin{align}\label{eq:8}
	\mathrm{P_c}(d|\mathbf{D}_d) = Q\left( \sqrt{2\gamma_{\Sigma}} \right),
\end{align}
where $\gamma_{\Sigma} = d_1\gamma_{SD} + d_2(\gamma_{SD}+\gamma_{Sel}) = d\gamma_{SD} + d_2 \gamma_{Sel}$ and $Q(x) = \frac{1}{\sqrt{2\pi}} \int_x^{+\infty} e^{-t^2/2}dt$ denotes the Q-function.

Taking into account the independence between $\gamma_{SD}$ and $\gamma_{Sel}$, we obtain the UPEP $\mathrm{P_u}(d|\mathbf{D}_d)$ given in Theorem~\ref{theorem:1} below.
%
%
\begin{theorem} \label{theorem:1}
Given the weight pattern $\mathbf{D}_d=\{d_1,d_2\}, d = d_1+d_2$, the UPEP $\mathrm{P_u}\left(d|\mathbf{D}_d\right)$ of PARC is given as follows:
\begin{align*}
\mathrm{P_u}\left(d|\mathbf{D}_d\right)\!=\!\left\{\!
{\begin{array}{*{20}{l}}
	\frac{1}{2}\left(1 - \sqrt{\frac{d\overline{\gamma}_{SD}}{1+d\overline{\gamma}_{SD}}}\right),\ \text{if}\ d_2 = 0  \\
	{\mathop \sum \limits_{j=1}^{N_r}} \Big(\!(-1)^{j-1}\!\mathop \sum \limits_{\begin{subarray}{c}
			{n_1}=1,\dots,{n_j} = 1 \\
			{n_1} \neq \dots \neq {n_j}
		\end{subarray}}^{N_r}
		\mathcal{I}_1\left(d\overline{\gamma}_{SD}, d_2\overline{\gamma}_{Sel,j}\right)\Big),\\
		~~\text{if}\ d_2 > 0
	\end{array}}
	\right.
\end{align*}
where \[
\mathcal{I}_1\left(a,b\right) = \frac{1}{2}\left(1 - \frac{a}{a-b}\sqrt{\frac{a}{a+1}} - \frac{b}{b-a}\sqrt{\frac{b}{b+1}}\right).
\]
\end{theorem}
%
\begin{IEEEproof}
See Appendix~\ref{App:A}. 
\end{IEEEproof}
Substituting UPEP in Theorem~\ref{theorem:1} into \eqref{eq:7} and \eqref{eq:5} we obtain the upper bound for the BER. Note that even though $d$ in \eqref{eq:5} can be as large as the codeword's length, i.e., $N$, the BER usually depends on few first values in fading channels. To give insightful understanding of PARC, we analyze the system diversity order.
\subsection{Finite-SNR Diversity Analysis}\label{sec:PARC_Diversity}
The classical definition of diversity order is defined as the negative exponent of the average BER as a function of SNR in log-log scale when the SNR tends to infinity \cite{Uysal2006}, which visually represents the slope of BER curve in infinity SNR domain. In this paper, since we are interested in finite SNR regime, we define the diversity order at a certain SNR $\gamma$ as follows:
\begin{eqnarray}\label{eq:InstDiversity}
\zeta(\gamma) \triangleq - \lim_{\Delta \rightarrow 1}\frac{\log[\mathrm{P}_2(\Delta\gamma)] - \log[\mathrm{P}_e(\gamma)]}{\log(\Delta\gamma) - \log(\gamma)},
\end{eqnarray}
which obviously matches the classical definition of diversity  when the SNR tends to infinity. Because the diversity order depends on the average SNR, we refer to $\zeta(\gamma)$ as \emph{instantaneous diversity order}. The key idea behind the definition is that it allows to study the behavior of the system at any SNR values. 

We might write $x \circeq \mathrm{SNR}^{-\eta}$ if $x$ has asymptotic diversity order $\eta$ (classical definition of diversity order). From \eqref{eq:5} we know that the diversity order of PARC is determined by $\mathrm{P_u}(d)$ because the input weight $w(d)$ of the channel code is constant. We first compute diversity order of the UPEP for a given weight pattern as in Theorem~\ref{theorem:2}.

\begin{theorem} \label{theorem:2}
Given the weight pattern $\mathbf{D}_d=\{d_1,d_2\}$ with $d = d_1+d_2$, the UPEP $\mathrm{P_u}\left(d|\mathbf{D}_d\right)$ in PARC is given as follows:
\begin{align*}
	\mathrm{P_u}\left(d|\mathbf{D}_d\right) \circeq \left\{
	{\begin{array}{*{20}{l}}
			\mathrm{SNR}^{-1},\ &\text{if}\ d_2 = 0 \\
			\mathrm{SNR}^{-(N_r+1)},\ &\text{if}\ d_2 > 0
		\end{array}}
		\right..
	\end{align*}
\end{theorem}
\begin{IEEEproof}
	See Appendix~\ref{App:B}.
\end{IEEEproof}
Theorem~\ref{theorem:2} states that $\mathrm{P_u}(d|\mathbf{D}_d)$ can have either diversity order one or diversity order $N_r+1$. From \eqref{eq:7} we conclude that $\mathrm{P_u}(d)$ is a combination of one factor with diversity of order one and one factor with diversity of order $N_r + 1$. The contribution of the factor with diversity order equal to one is computed from \eqref{eq:9} as follows:
\begin{align*}
	p(D_1\triangleq \{d,0\}) =  \frac{\mathcal{C}^{N/2}_d}{\mathcal{C}^N_d} = \prod_{k=0}^{d-1} \frac{N-2k}{2N - 2k}.
\end{align*}
In practical systems, the codeword length $N$ is usually much larger than $d$, then $p(D_1)$ can be well-approximated as
\begin{align} \label{eq:9}
	p(D_1) \simeq \left(\frac{1}{2}\right)^d \leq \left(\frac{1}{2}\right)^f.
\end{align}
Substituting the result in Theorem~\ref{theorem:2} into \eqref{eq:5} we can write:
\begin{align}\label{eq:10}
	\mathrm{Pe} \circeq K 2^{-f} \mathrm{SNR}^{-1} + \mathrm{SNR}^{-N_r-1},
\end{align}
where $K$ is the normalized constant that depends on the channel code and network topology. 

From \eqref{eq:InstDiversity} and \eqref{eq:10} we conclude that the instantaneous diversity order of PARC consists of one factor which achieves full diversity order and one factor which achieves a diversity of order one. The impact of the diversity order one factor is inversely proportional to the channel code' strength, i.e., its minimum distance. For a strong code with large minimum distance $f$, the impact of the diversity one factor is negligible compared with the diversity order $N_r+1$ factor, resulting the system can achieve full diversity $N_r+1$ in the finite SNR regime. This result is important because the operating SNR regime in practical systems is usually finite. 
\section{Performance Analysis for Network Coding based Cooperation} \label{sec:PerformanceNCC}

In this section, we analyze the BER and diversity order of NCC. Using the equivalent channel, the two-hop network-coded signal can be modeled as if it was conveyed by a single channel whose instantaneous SNR is $\gamma_{\rm {NC}}$ \cite{Nasri2013}. 
\subsection{Derivation of BER}
Recalling that in NCC, the destination applies the BCJR algorithm on the compound code $\mathbf{G}$, which is described in Section~\ref{sec:NCC}. The compound code $\mathbf{G}$ has compound input $\mathbf{X} = [\mathbf{x}_1, \mathbf{x}_2, \mathbf{x}_{NC}]$ and channel output $\mathbf{Y} = [\mathbf{y}_{S_1D}, \mathbf{y}_{S_2D}, \mathbf{y}_{R_{NC}D}]$. Note that the output of $\mathbf{G}$ undergoes some  block fading channels with three blocks $\gamma_{S_1D}, \gamma_{S_2D}$, and $\gamma_{NC}$, and it decodes the data messages instantaneously. Consider $\mathbf{G}$ as a regular channel code, the BER of source $\mathrm{S}_i$ is calculated as:
\begin{align}\label{eq:Pe_NC}
\widetilde{\mathrm{Pe}}_i = \frac{1}{2}\mathop \sum \limits_{d = F}^{ + \infty } w_i(d) \mathrm{\widetilde{P}_u}(d),~ i = 1,2,
\end{align}
where $F$ is the minimum distance of the compound code $\mathbf{G}$, $w_i(d)$ denotes input weights corresponding to source $\mathrm{S}_i$ in the compound codeword, and $\mathrm{\widetilde{P}_u}(d)$ is the UPEP of receiving a super codeword with the output weight $d$, assuming that the all-zero compound codeword has been transmitted ($\mathbf{c}_1 = \mathbf{c}_2 = \mathbf{0}$). To derive \eqref{eq:Pe_NC}, it requires to know the minimum distance $F$ of the compound code, the input weight $w_i(d)$ and how $d$ bits in the compound codeword $\mathbf{X}$ are distributed among three channels $\mathrm{S}_1 \to \mathrm{D}$, $\mathrm{S}_2 \to \mathrm{D}$, and $\mathrm{R}_{NC} \to \mathrm{D}$. Denote $\mathbf{W}_d = \{d_1, d_2, d_R\}$  as the weight pattern that specifies how $d$ weights are distributed among these channels, where $d_i$ is the output weight of the individual codeword transmitted via the channel $\mathrm{S}_i \to \mathrm{D}$ or channel $\mathrm{R}_{NC} \to \mathrm{D}$. By definition, $d = d_1+d_2+ d_R$. The input weight and the pattern can be computed via heuristic searching of the trellis of $\mathbf{G}$. The following result is important for further analysis.
%
\begin{lemma}\label{lemma:1}
	The minimum distance $F$ of the compound code $\mathbf{G}$ is equal to twice the minimum distance of the single code $\mathbf{g}$, $F=2f$, and the weight pattern $\mathbf{W}_F$ has one of the following values $\{f,f,0\},\ \{f,0,f\},\ \{0,f,f\}$.
\end{lemma}
%
\begin{lemma}\label{lemma:2}
	For any pattern $\mathbf{W}_d=\{d_1, d_2, d_R\}$ of the compound codeword $\mathbf{X}$ with output weight $d > F$, there are at least two non-zero elements in $\mathbf{W}_d$.
\end{lemma}
%
The proof of Lemma~\ref{lemma:1} and Lemma~\ref{lemma:2} are given in \cite{Vu2013}. 
%
Lemma~\ref{lemma:1} and Lemma~\ref{lemma:2} provide an important information about the output weights of the compound code: $d$ weights of the compound code always experience at least two independent channels. Furthermore, the number of patterns is finite and strictly defined by $\mathbf{G}$. 

Using Lemma~\ref{lemma:1} and \ref{lemma:2} we can reformulate \eqref{eq:Pe_NC} as follows:
\begin{align}\label{eq:Pe_NC1}
\widetilde{\mathrm{Pe}}_i = \frac{1}{2}\mathop \sum \limits_{d = F}^{ + \infty } \sum\limits_{\mathbf{W}_d} w_i\left(\mathbf{W}_d\right) \mathrm{\widetilde{P}_u}\left( d|\mathbf{W}_d\right),
\end{align}
where $\mathrm{\widetilde{P}_u}(d|\mathbf{W}_d)$ is the UPEP depending on the pattern $\mathbf{W}_d$ and is the expectation of the CPEP over the fading channels:
\begin{align*}
	\mathrm{\widetilde{P}_u}(d|\mathbf{W}_d) = \mathbb{E}\{\mathrm{\widetilde{P}_c}(d|\mathbf{W}_d)\}.
\end{align*}
It is assumed that the erroneous detected symbol could only be one of the nearest neighbor symbols. Using the Gray mapping, each closest symbol error only causes one coded bit error. Therefore, the CPEP $\mathrm{P_c}(d|\mathbf{D}_d)$ is approximated as \cite{Vu2013}:
\begin{align}\label{eq:15}
\mathrm{\widetilde{P}_c}\left(d|\mathbf{W}_d\right) = Q\left(\sqrt{2\gamma_{\Sigma_{NC}}}\right),
\end{align}
where $\gamma_{\Sigma_{NC}} = d_1\gamma_{S_1D} + d_2\gamma_{S_2D} + d_R\gamma_{NC}$ is defined as the total SNR at the destination in NCC. Because the three channels in $\gamma_{\Sigma_{NC}}$ are mutually independent, the MGF of the total SNR can be computed as follows:
\begin{align*}
\Psi_{\gamma_{\Sigma_{NC}}}(s) = \Psi_{\gamma_{S_1D}}(d_1s) \times \Psi_{\gamma_{S_2D}}(d_2s) \times \Psi_{\gamma_{NC}}(d_Rs).
\end{align*}
Applying the MGF method \cite{Simon2005} we can derive the UPEP $\mathrm{\widetilde{P}_u}\left(d|\mathbf{W}_d\right)$ in NCC as in Theorem~\ref{theorem:3}.
%
\begin{theorem} \label{theorem:3}
	Given the weight pattern $\mathbf{W}_d=\{d_1,d_2,d_R\}$, $d = d_1+d_2+d_R$, the UPEP $\mathrm{\widetilde{P}_u}\left(d|\mathbf{W}_d\right)$ of the compound code in NCC has a form given by:
	\begin{align}
	\mathrm{\widetilde{P}_u}\left(d|\mathbf{W}_d\right) = \left\{
	{\begin{array}{*{20}{l}}
		\mathcal{I}_1\left(d_1\overline{\gamma}_{S_1D}, d_2\overline{\gamma}_{S_2D}\right),\ &\text{if}\ d_R = 0  \\
		{\mathop \sum \limits_{j=1}^L} (-1)^{j-1} \Omega_1, &\text{if}\ d_1 = 0  \\
		{\mathop \sum \limits_{j=1}^L} (-1)^{j-1} \Omega_2, &\text{if}\ d_2 = 0 \\
		{\mathop \sum \limits_{j=1}^L} (-1)^{j-1} \Omega_3, &\text{if}\ d_1d_2d_R \neq 0
		\end{array}}
	\right.,\notag
	\end{align}
	where $\mathcal{I}_1(a,b)$ has been defined in Theorem~\ref{theorem:1}, 
	\begin{align*}
	\Omega_1 &= \mathop \sum \limits_{\begin{subarray}{c}
				{n_1}=1,\dots,{n_j} = 1 \\
				{n_1} \neq \dots \neq {n_j}
				\end{subarray}}^L
			\mathcal{I}_1\left(d_2\overline{\gamma}_{S_2D}, d_R\overline{\gamma}_{NC,j}\right), \\
	\Omega_2 &= \mathop \sum \limits_{\begin{subarray}{c}
				{n_1}=1,\dots,{n_j} = 1 \\
				{n_1} \neq \dots \neq {n_j}
				\end{subarray}}^L
			\mathcal{I}_1\left(d_1\overline{\gamma}_{S_1D}, d_R\overline{\gamma}_{NC,j}\right),\\
	\Omega_3 &= \mathop \sum \limits_{\begin{subarray}{c}
				{n_1}=1,\dots,{n_j} = 1 \\
				{n_1} \neq \dots \neq {n_j}
				\end{subarray}}^L
			\mathcal{I}_2\left(d_1\overline{\gamma}_{S_1D}, d_2\overline{\gamma}_{S_2D}, d_R\overline{\gamma}_{NC,j}\right),
	\end{align*}
and
	\begin{align}	
\mathcal{I}_2\Big(a,b,c\Big) &= \frac{1}{2}\Big(1 - \frac{a^2}{(a-b)(a-c)}\sqrt{\frac{a}{a+1}} - \notag \\
        &\frac{b^2}{(b-a)(b-c)}\sqrt{\frac{b}{b+1}} - \frac{c^2}{(c-a)(c-b)}\sqrt{\frac{c}{c+1}}\Big).\notag
	\end{align}
\end{theorem}
\begin{IEEEproof}
	See Appendix ~\ref{App:C}.
\end{IEEEproof}

It is worthnoting in \eqref{eq:Pe_NC1} that the BER of each source in NCC is a sum of terms given in Theorem~\ref{theorem:3}, weighted by their corresponding input weights ${\mathrm{w}_i}\left( {\mathbf{W}_d} \right)$. In NCC, the weight pattern only holds a few values and the input weights are computed from the extended distance spectrum. Table~\ref{table:1} gives an example of the distance spectrum of the compound code.
\begin{table}[ht]
	\caption{Input weight and output weight distribution at $d = F =24$ of compound code $\mathbf{G}$ in \eqref{eq:G}, $\mathbf{g}=[23,\ 35,\ 37]$} 
	\centering 
	\begin{tabular}{c c c c c} 
		\hline\hline 
		$w_1$ & $w_2$ & $d_1$ & $d_2$ & $d_R$ \\ [0.5ex] 
		\hline 
		0 & 12 & 0 & 12 & 12\\
		12 & 0 & 12 & 0 & 12\\
		12 & 12 & 12 & 12 & 0\\ [1ex] 
		\hline 
	\end{tabular}	\label{table:1} 
	\vspace{-0.2cm}
\end{table}
\subsection{Diversity analysis}
Since the BER in NCC is a linearly proportional to the UPEP $\mathrm{\widetilde{P}_u}\left(d|\mathbf{W}_d\right)$ via the input weights, the diversity order of NCC is equal to diversity order of the UPEP. Employing asymptotic equivalent notation as in the previous section, the diversity order of the UPEP is given as the below theorem.
%
\begin{theorem} \label{theorem:4}
    Given the weight pattern $\mathbf{W}_d=\{d_1,d_2,d_R\}$ with $d = d_1+d_2+d_R$, the UPEP $\mathrm{\widetilde{P}_u}\left(d|\mathbf{W}_d\right)$ in NCC has an asymptotic equivalent form as follows:
    \begin{align*}
        \mathrm{\widetilde{P}_u}\left(d|\mathbf{W}_d\right) \circeq \left\{
            {\begin{array}{*{20}{l}}
            \mathrm{SNR}^{-2},\ &\text{if}\ d_R = 0  \\
            \mathrm{SNR}^{-(N_r+1)},\ &\text{if}\ d_1 = 0\ \text{or}\ d_2 = 0 \\
            \mathrm{SNR}^{-(N_r+2)}, &\text{if}\ d_1d_2d_R \neq 0
            \end{array}}
        \right..
    \end{align*}
\end{theorem}
The proof of Theorem~\ref{theorem:4} is given in Appendix~\ref{App:B}.

It is shown from \eqref{eq:Pe_NC1} and Theorem~\ref{theorem:4} that the BER in NCC is a combination of three factors whose diversity orders are respectively $2,\ N_r+1$ and $N_r+2$. As the contribution of these factors are comparable and equal input weights of the compound code (shown in Table~\ref{table:1} as an example), the diversity order of NCC is dominated by the diversity order 2 factor. Consequently, NCC achieves diversity order 2 regardless the channel code and the total number of available relays.
\section{Numerical Results}\label{sec:Results}
\begin{figure*}[!t]
	\normalsize
	\centering
	\subfigure[$N_r = 2$]{\includegraphics[width = 0.49\textwidth]{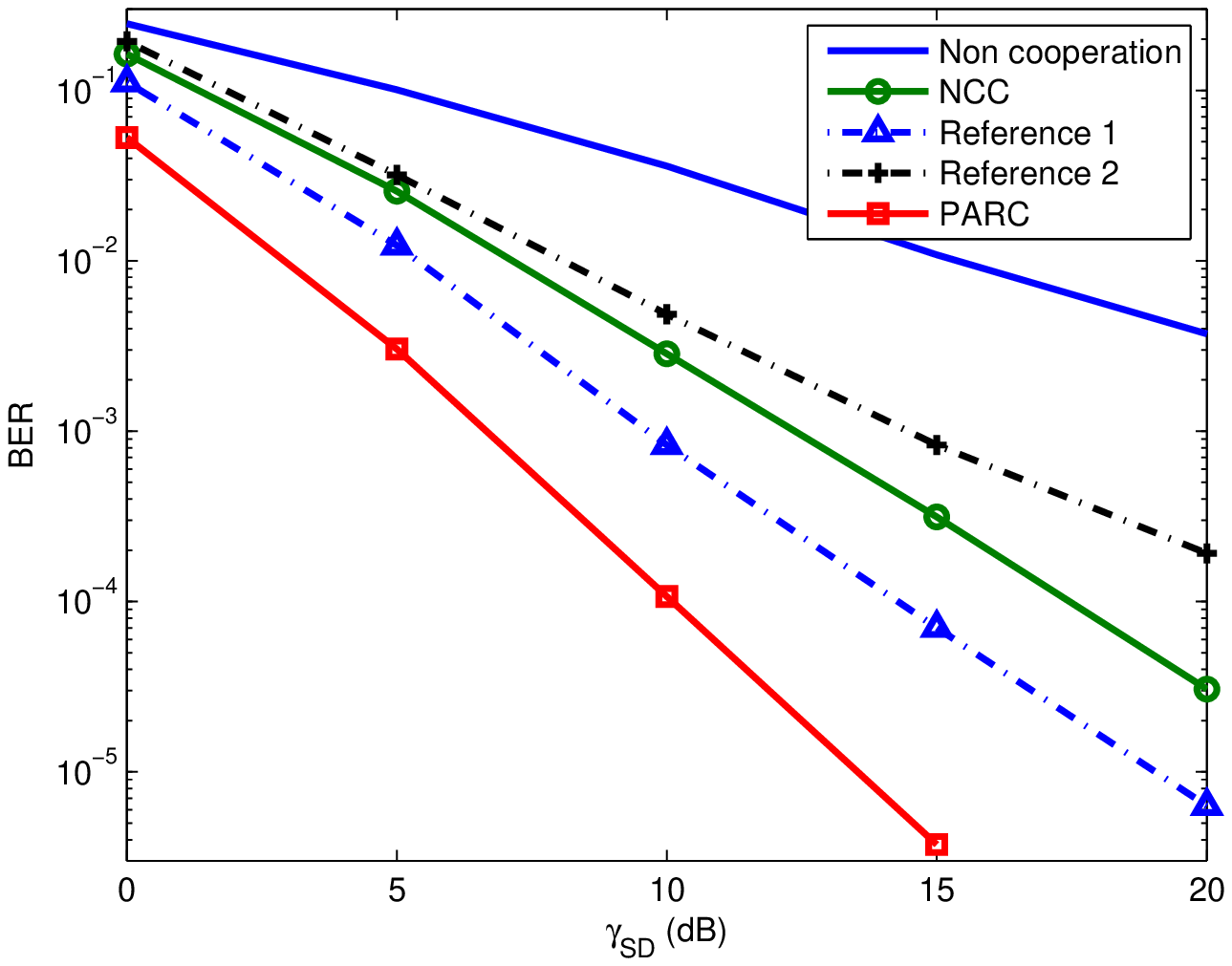}}
	\subfigure[$N_r = 3$]{\includegraphics[width = 0.49\textwidth]{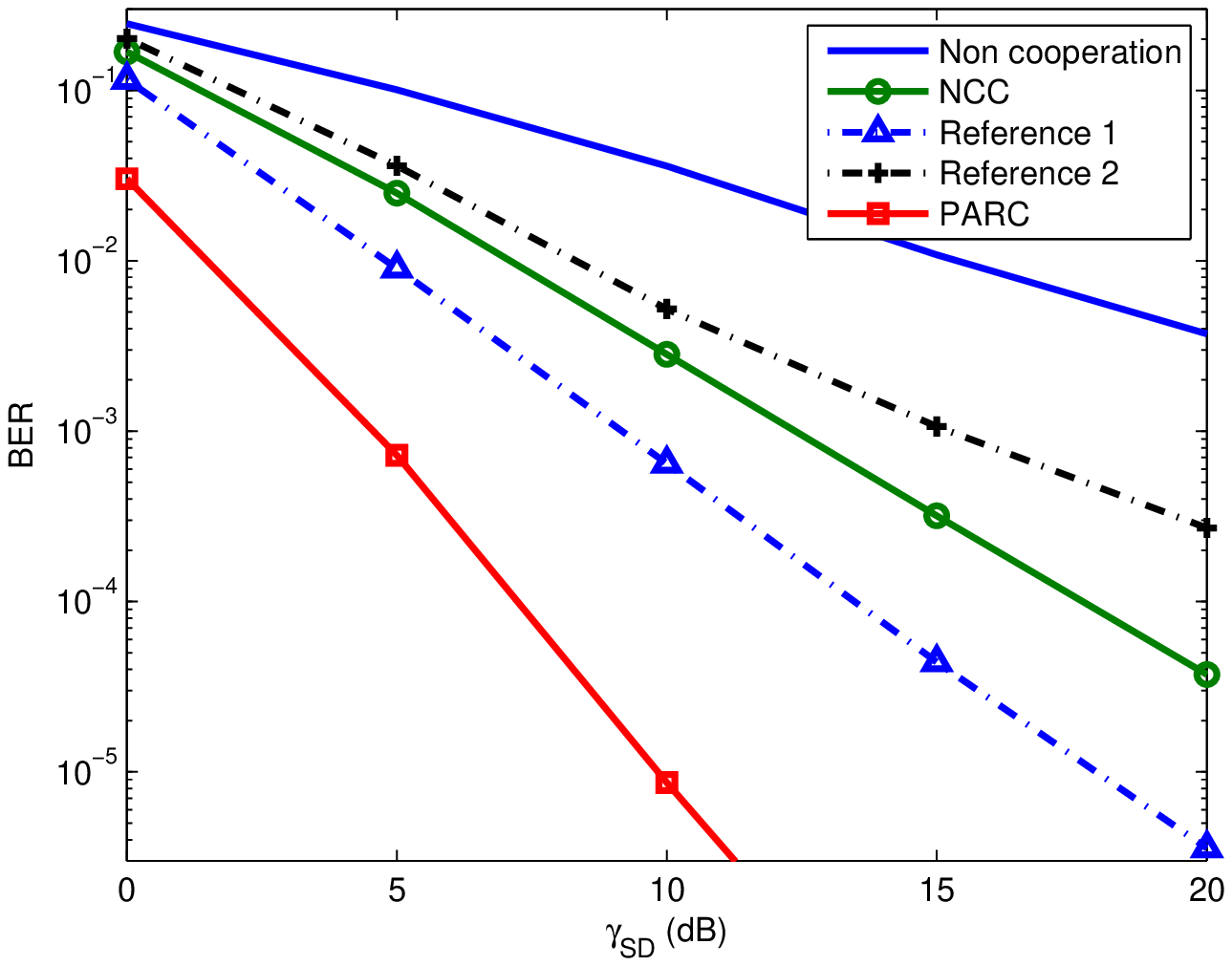}}
	\caption{Performance comparison between PARC and NCC when the CC [133 165 171] with the minimum distance $f = 15$ and the rate 1/3 is used.}
	\label{fig:CC133_165_171}
\end{figure*}
\begin{figure*}[!t]
	\normalsize
	\centering
	\subfigure[$N_r = 2$]{\includegraphics[width = 0.49\textwidth]{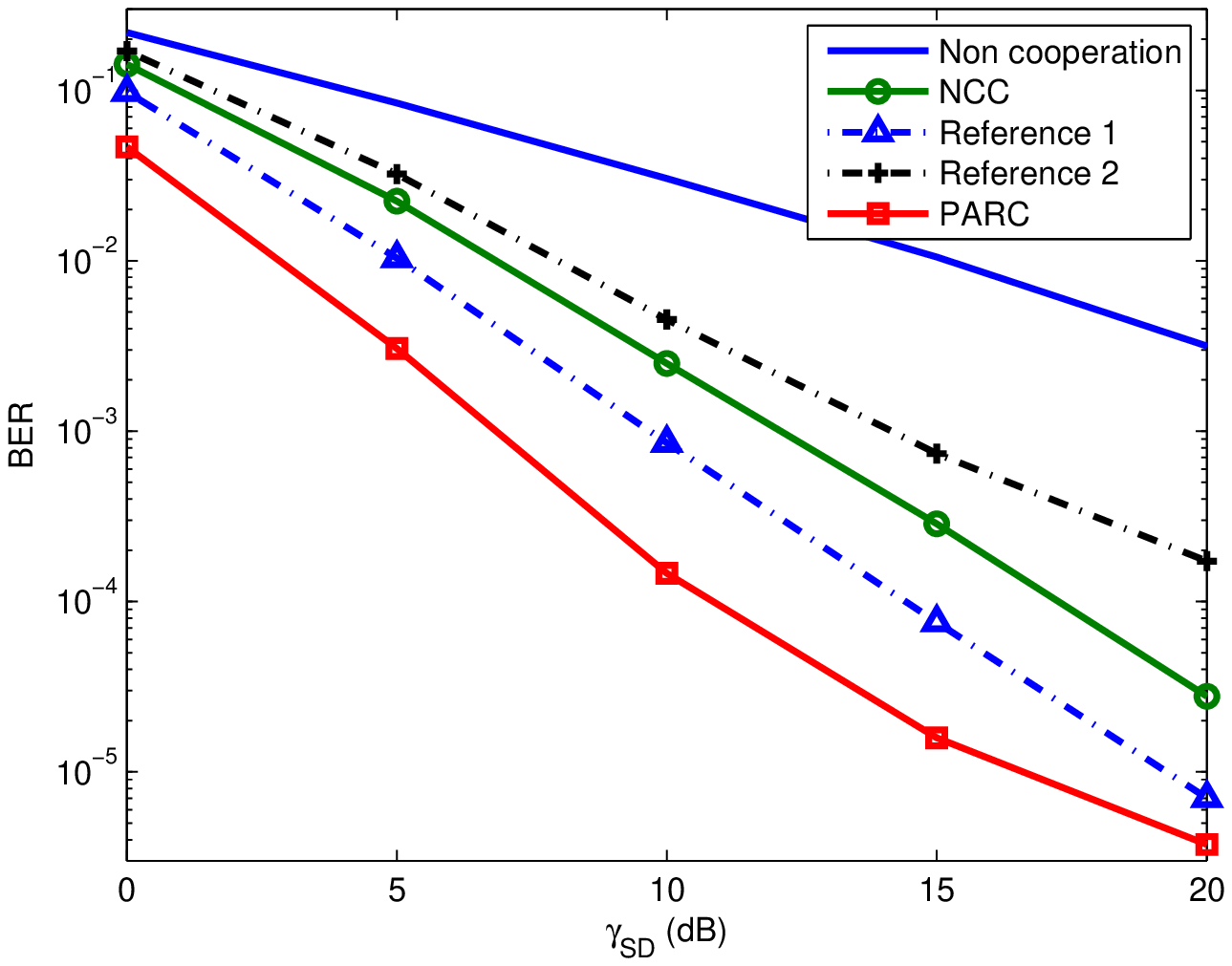}}
	\subfigure[$N_r = 3$]{\includegraphics[width = 0.49\textwidth]{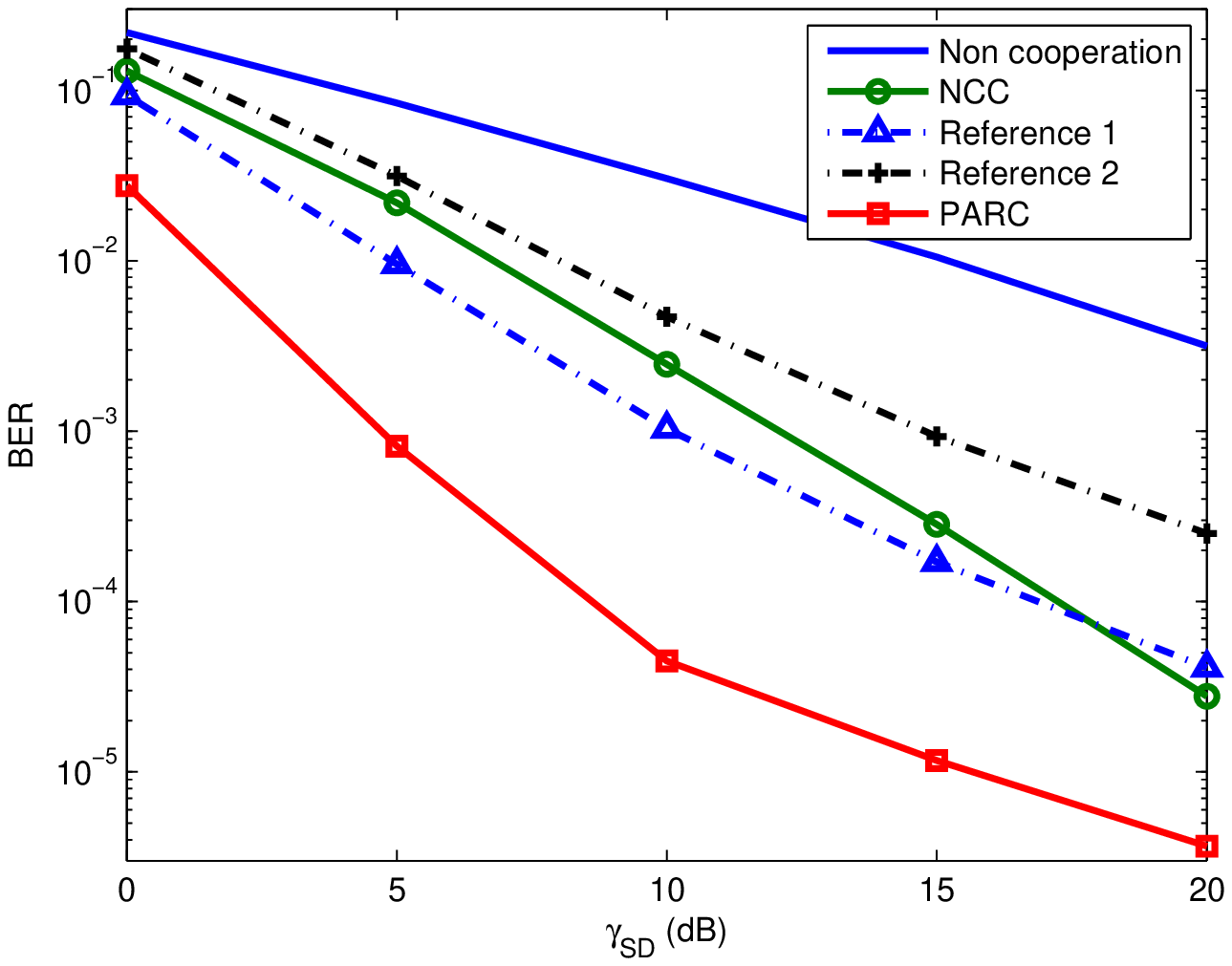}}
	\caption{Performance comparison between PARC and NCC when the CC [25 33 37] with the minimum distance $f = 12$ and the rate 1/3 is used.}
	\label{fig:CC25_33_37}
\end{figure*}
This section shows simulation results to confirm the effectiveness of the proposed system described in Section~\ref{sec:SystemModel}. All channels are subject to quasi-static block Rayleigh fading plus AWGN. Because we focus on the diversity order, and the modulation order does not change the system diversity order, BPSK modulation and binary network coding are carried out in simulations. The data packet length is equal to $1024$bits. We consider symmetric network, i.e., $\overline{\gamma}_{S_iR_j} = \overline{\gamma}_{SR}, \overline{\gamma}_{R_jD} = \overline{\gamma}_{RD}, \overline{\gamma}_{S_iD} = \overline{\gamma}_{SD}, \forall i,j$. The relays locate at the middle of the sources and the destination and the pathloss exponent is 3.5, resulting in that the average SNR in source-relay channels and relay-destination channels are 10.5dB better than source-destination channels. Note that our analysis validates for arbitrary locations of the relays. The channel code is chosen as the one that optimizes both the minimum distance and distance spectrum in block Rayleigh fading channels \cite{Frenger1999}. Different channel codes $\mathbf{g}$ are compared.

We also present the performance of two reference schemes. The first reference scheme (named \emph{Reference 1} in the figures) is based on fractional repetition coding cooperation \cite{Eckford2008,Khormuji2009}. The second reference scheme employs factional repetition coding together with network coding (named \emph{Reference 2} in the figures). All relays are active and share the relaying phase in two reference schemes. In \emph{Reference 1}, since the relays help the sources separately, each relay forwards $1/(2N_r)$ of the estimated codeword. In \emph{Reference 2}, all relays use NC to help the sources and each relay randomly forwards $1/N_r$ of the network-coded codeword. We note that no relay selection is used in the reference schemes.

Figure~\ref{fig:CC133_165_171} compares the performance of PARC and referenced schemes for the channel code [133 165 171] with code rate 1/3 and the minimum distance $f=15$. The total number of relays $N_r$ equal to 2 and 3 are plotted. The observed performance region satisfies $\mathrm{BER} \geq 10^{-6}$ because this is the target BER for most practical applications. It is shown in the figure that the proposed PARC achieves an instantaneous diversity order of 3 and 4 (full diversity order) in the observing SNR range when the total number of relays is 2 and 3, respectively. Such expected result can be explained from Theorem~\ref{theorem:2} that in this case, the impact of the diversity one factor equals $p(\mathrm{D}_1) = (1/2)^{f} \simeq 3.10^{-5}$, which is negligible. Therefore, the diversity order of PARC is determined by the full diversity factor in the observing SNR region. In contrast, NCC always achieves a diversity of order 2, which is a consequence of Theorem~\ref{theorem:4}. A huge SNR gain is achieved by PARC. In particular, PARC outperforms all other schemes about 5dB for $N_r = 2$ and 9dB for $N_r = 3$ at BER $= 10^{-4}$. Another observation is that Reference 1 also surpasses NCC because the relayed symbols in Reference 1 see more spatial diversity gain than that in NCC. When SNR tends to infinity, NCC may outperform PARC because the diversity order of PARC will collapse to one while NCC still has diversity order equal to 2. From the practical system point of view, this crossing-point might not weaken the advantage of PARC over NCC since practical systems usually operate at finite SNR regime.

Figure~\ref{fig:CC25_33_37} shows the performance comparisons when the channel code [25 33 37] with rate 1/3 is used. The minimum distance of this code is equal to 12. It is not surprised that NCC always achieves diversity order 2 for both $N_r = 2$ or $N_r = 3$ and the performance of NCC in both cases is similar. It is observed that PARC only achieves full diversity order in low SNRs. More specifically, PARC achieves diversity order 3 in the SNR range until 10dB when $N_r=2$ and diversity order 4 until SNR = 5dB when $N_r = 3$. When SNR increases, a degradation in instantaneous diversity order is observed, which is predicted by our analysis (for this code, the contribution of diversity order one factor approximately is $2^{-f} \simeq 2.4e-4$). However, a similar SNR gain as for strong code [133 165 171] is achieved by PARC at BER of $10^{-4}$, which is about 5dB for $N_r=2$ and 7dB for $N_r = 3$. A sound interesting observation is that the performance Reference 1 for $N_r = 3$ is worse than that for $N_r = 2$. This is because in the later, the relay forwards less symbols in $N_r=3$ than in $N_r=2$ case, and the channel code is not strong enough to compensate for less relayed symbols in $N_r = 3$ case \cite{Vu2013a}.  

\begin{figure}[!t]
	\centering
	\includegraphics[width = \columnwidth]{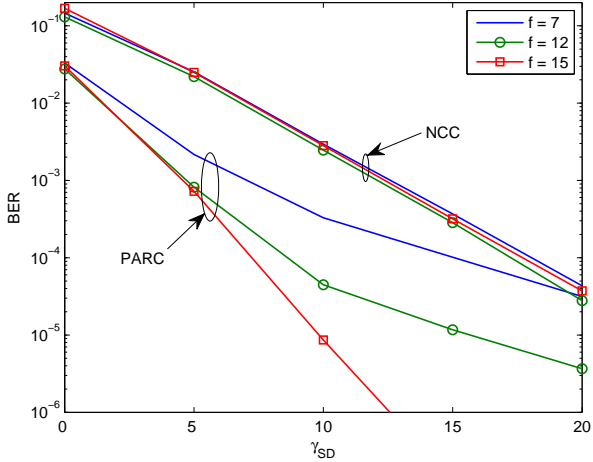}
	\caption{Performance comparison between PARC and NCC for different minimum distances, which corresponds to code's correction capacity. Three codes with rate 1/3 are compared: CC [5 7 5] with $f = 7$, CC [25 33 37] with $f = 12 $, and CC [133 165 171] with $f = 15$.}\label{fig:CodesCompare}
	\vspace{-0.2cm}
\end{figure}

Figure~\ref{fig:CodesCompare} compares the performance between PARC and NCC for different channel codes and $N_r = 3$. Three codes with different error correction capabilities are presented: the weak code [5 7 5] with small minimum distance $f = 7$, the moderate code [25 33 37] with $f = 12$, and the strong code [133 165 171] with $f = 15$. Full diversity order is observed in low SNRs for all codes. When SNR increases only the strong code achieve full diversity order. The weak code starts losing diversity order earliest at SNR of 5dB, while the moderate code's diversity degrades at 10dB. Compared with NCC, however, PARC significantly outperforms for all codes in the observed SNRs. 

In conclusion, the most effectiveness of the proposed PARC is that it can achieve full (instantaneous) diversity order in the low and medium SNR regime, which in turn results in a large SNR gain in finite SNR region. This is crucial for practical systems because their operating SNRs is usually finite. 

\section{Conclusions and Discussions}\label{sec:Conclusions}
\balance
We have proposed a novel cooperative scheme for a two-source multiple-relay networks that combines relay selection and partial relaying to effectively exploit the spatial diversity. In the proposed scheme, the selected relay retransmits half of the estimated codeword in order to satisfy the spectral efficiency constraint. We have analytically shown that the proposed scheme can gain full diversity order in finite SNR regime (instantaneous diversity) when a suitable channel code is used. It has been shown that the instantaneous diversity order is a function of the minimum distance of the code and the operating SNR. Numerical results show a significant SNR improvement the proposed scheme compared with reference schemes. 

The proposed partial relaying cooperation can easily be extended to general networks that consist of multiple sources and a given finite time slots for relaying. In this case, the selected relays might forward a number of symbols different from half of codeword. A promising application of PARC is to design for the sources which have different error correction capacities to achieve a given target BER. As such, how many relayed symbols for a source should be carefully chosen depending on the strength of its channel code. 
\appendices
\section{Proof of Theorem~\ref{theorem:1}}
\label{App:A}
\renewcommand{\theequation}{A.\arabic{equation}}
\setcounter{equation}{0}
Because the relayed symbols are randomly distributed on the codeword, the weight $d_2$ on the relayed block can have any integer value in $[0,d]$. Denote $\mathrm{D}_1 = \{d,0\}$ as the weight pattern in which all $d$ weights are not relayed. Then the weight pattern in general has one of two forms, $\mathrm{D}_1 = \{d,0\}$ and $\mathbf{D}_d \neq \mathrm{D}_1$. Using the MGF, the UPEP can be computed for general cases as follows:
	\begin{align*}
	\mathrm{P_u}(d|\mathbf{D}_d) = \frac{1}{\pi}\int_{0}^{\pi/2} \Psi_{\gamma_{\Sigma}} \left(\frac{1}{\sin\theta^2}\right)d\theta.
	\end{align*}
\begin{itemize}
	\item \textbf{Case 1}: $\mathbf{D}_d = D_1$. In this case, all $d$ weights locate in the source-destination block, resulting in $\gamma_{\Sigma} = d\gamma_{SD}$ and $\Psi_{\gamma_{\Sigma}}(s) = \Psi_{\gamma_{SD}}\left(ds\right)$. In this case we have:
	\begin{align}\label{eq:A1}
	\mathrm{P_u}(d|\mathrm{D}_1) &= \frac{1}{\pi}\int_{0}^{\pi/2} \frac{\sin\theta^2}{\sin\theta^2 + d\overline{\gamma}_{SD}}d\theta \notag \\
	&= \frac{1}{2}\left(1 - \sqrt{\frac{d\overline{\gamma}_{SD}}{1 + d\overline{\gamma}_{SD}}}\right).
	\end{align}
	\item \textbf{Case 2}: $\mathbf{D}_d \neq \mathrm{D}_1$. There is always $d_2$ weights are relayed, resulting in $\Psi_{\gamma_{\Sigma}}(s) = \Psi_{\gamma_{SD}}\left(ds\right)\times \Psi_{\gamma_{\Sigma}}\left(d_2s\right)$. From \eqref{eq:4} we have:
	\begin{align} \label{eq:A2}
	\mathrm{P_u}&\left(d|\mathbf{D}_d\right)	= {\mathop \sum \limits_{j=1}^{N_r}}  \Big( (-1)^{j-1} \mathop \sum \limits_{\begin{subarray}{c}
		{n_1}=1,\dots,{n_j} = 1 \\
		{n_1} \neq \dots \neq {n_j}
		\end{subarray}} ^{N_r} 	\Big. \notag \\
	 &\Big. \frac{1}{\pi} \int \limits_{0}^{\pi/2}\frac{\sin\theta^4}{\left(\sin\theta^2 + d\overline{\gamma}_{SD}\right)\left(\sin\theta^2 + d_2\overline{\gamma}_{Sel, j}\right)}d\theta \Big) \notag \\
	=& {\mathop \sum \limits_{j=1}^{N_r}} \Big( (-1)^{j-1} \mathop \sum \limits_{\begin{subarray}{c}
		{n_1}=1,\dots,{n_j} = 1 \\
		{n_1} \neq \dots \neq {n_j}
		\end{subarray}} ^{N_r} \mathcal{I}_1\left(d\overline{\gamma}_{SD}, d_2\overline{\gamma}_{Sel,j}\right)\Big),
	\end{align}
	where
	\[\mathcal{I}_1(a,b) = \frac{1}{2}\left(1 - \frac{a}{a-b}\sqrt{\frac{a}{1+a}} -  \frac{b}{b-a}\sqrt{\frac{b}{1+b}}\right).\]
\end{itemize}
\section{Proof of Theorem~\ref{theorem:2}}
\label{App:B}
\renewcommand{\theequation}{B.\arabic{equation}}
\setcounter{equation}{0}
The diversity order is defined as the negative exponent of UPEP in log-scale when the average SNR $\overline{\gamma}$ tends to infinity
\begin{align}\label{eq:B1}
\tau_d = - \lim_{\overline{\gamma} \rightarrow \infty} \left(\frac{\log \mathrm{P_u}\left(d|\mathbf{D}_d\right)}{\log \overline{\gamma}}\right).
\end{align}
Using the upper bound of UPEP \cite{Simon2005} as $\mathrm{P_u}\left(d|\mathbf{D}_d\right) \leq \frac{1}{2}\Psi_{\gamma_{\Sigma}}(1/2) < \Psi_{\gamma_{\Sigma}}(1/2)$ and recall \eqref{eq:B1} we have
\begin{align}\label{eq:B2}
\tau_d \geq - \lim_{\overline{\gamma} \rightarrow \infty} \left(\frac{\log \Psi_{\gamma_{\Sigma}}(1/2)}{\log \overline{\gamma}}\right).
\end{align}
Similar to Appendix~\ref{App:A}, we consider two cases.

\begin{itemize}
	\item \textbf{Case 1}: $\mathbf{D}_d = \mathrm{D}_1$. There is not any symbol helped by the relay and $\Psi_{\gamma_{\Sigma}}\left(1/2\right) = \Psi_{\gamma_{SD}}\left(d/2\right)$. The diversity order in this case is given by
	\begin{align}\label{eq:B3}
	\tau_d &\geq -\lim_{\overline{\gamma} \rightarrow \infty} \frac{\log \Psi_{\gamma_{SD}}\left(d/2\right)}{\log \overline{\gamma}} \notag \\
	& \geq -\lim_{\overline{\gamma} \rightarrow \infty}\left(\frac{\left(1 + d\overline{\gamma}_{SD}/2\right)^{-1}} {\log \overline{\gamma}}\right) = 1. 
	\end{align}
	Then the UPEP has diversity order of 1 when $d_2 = 0$ and we can write $\mathrm{P_u}\left(d|\mathrm{D}_1\right) \circeq \mathrm{SNR}^{-1}$.
	
	\item \textbf{Case 2}: $\mathbf{D}_d \neq \mathrm{D}_1$. The MGF of $\gamma_{\Sigma}$ in this case has a form of $\Psi_{\gamma_{\Sigma}}\left(1/2\right) = \Psi_{\gamma_{SD}}\left(d/2\right)\times \Psi_{\gamma_{Sel}}\left(d_2/2\right)$. Consequently, the diversity order is given as follows:
	\begin{align}\label{eq:B4}
	\tau_d &\geq -\lim_{\overline{\gamma} \rightarrow \infty} \frac{\log \Psi_{\gamma_{SD}}\left(d/2\right)}{\log \overline{\gamma}} \underbrace {-\lim_{\overline{\gamma} \rightarrow \infty} \frac{\log \Psi_{\gamma_{Sel}}\left(d_2/2\right)}{\log \overline{\gamma}}}_{\tau_{Sel}} \notag \\
	&= 1 +  \tau_{Sel},
	\end{align}
	where $\tau_{Sel}$ is the diversity order of the best relay signal (without direct link). It has shown in \cite{Bletsas2006} that the best relay selection achieves diversity order is equal to the total number of available relays, we have $\tau_{Sel} = N_r$. Therefore the system diversity order in this case is equal to $N_r + 1$. In order words, we can write $\mathrm{P_u}\left(d|\mathbf{D}_d \neq \mathrm{D}_1\right) \circeq \mathrm{SNR}^{-(N_r+1)}$.
\end{itemize}
Combine the two cases above we complete the proof of Theorem~\ref{theorem:2}.
\section{Proof of Theorem~\ref{theorem:3}}
\label{App:C}
\renewcommand{\theequation}{C.\arabic{equation}}
\setcounter{equation}{0}

It is noted that the total SNR in NCC is given by $\gamma_{\Sigma_{NC}} = d_1\gamma_{S_1D} + d_2\gamma_{S_2D} + d_R\gamma_{NC}$. To derive the UPEP for NCC, the MGF method is employed. Lemma~\ref{lemma:1} and Lemma~\ref{lemma:2} state that there are at least two weights in $\{d_1, d_2, d_R\}$ are non-zero. Therefore, the weight pattern $\mathbf{W}_d$ can only has one of these four cases: 1) $d_R = 0$, 2) $d_1 = 0$, 3) $d_2 = 0$ and 4) $d_1d_2d_R \neq 0$. Consequently, the total SNR $\gamma_{\Sigma_{NC}}$ has one of corresponding four values.

\begin{itemize}
\item \textbf{Case 1}: $d_R$ = 0, there is not any weight on the relay channel. The total SNR has a form of $\gamma_{\Sigma_{NC}} = d_1\gamma_{S_1D} + d_2 \gamma_{S_2D}$, and its MGF is given as follows:
    \begin{align}\label{eq:40}
        \Psi_{\gamma_{\Sigma_{NC}}}(s) &= \Psi_{\gamma_{S_1D}}(d_1s) \times \Psi_{\gamma_{S_2D}}(d_2s) \notag \\
        &= \frac{1}{1 - d_1\overline{\gamma}_{S_1D} s}\frac{1}{1 - d_2\overline{\gamma}_{S_2D} s}.
    \end{align}
    Now the UPEP $\mathrm{\widetilde{P}_u}\left(d|\mathbf{W}_d\right)$ can be computed using the MGF method \cite{Simon2005} as follows:
    \begin{align} \label{eq:41}
        \mathrm{\widetilde{P}_u}\left(d|\mathbf{W}_d\right)&=\frac{1}{\pi}\int \limits_{0}^{\pi/2} \Psi_{\gamma_{\Sigma_{NC}}}\left(\frac{1}{\sin\theta^2}\right)d\theta \notag \\
        & = \mathcal{I}_1\left(d_1\overline{\gamma}_{S_1D},d_2\overline{\gamma}_{S_2D}\right),
    \end{align}
    where $\mathcal{I}_1\left(a,b\right)$ has been defined in Appendix~\ref{App:A}.

\item \textbf{Case 2}: $d_1 = 0$. In this case, the total SNR equals $\gamma_{\Sigma_{NC}} = d_2\gamma_{S_2D} + d_R \gamma_{NC}$. Given the PDF of $\gamma_{NC}$ in~\eqref{eq:6}, the MGF of the total SNR is given as follows:
    \begin{align}\label{eq:43}
        &\Psi_{\gamma_{\Sigma_{NC}}}(s) = \Psi_{\gamma_{S_2D}}(d_2s) \times \Psi_{\gamma_{{NC}}}(d_Rs)\notag \hfill \\
         &= \mathop \sum \limits_{j=1}^{N_r} \!\Big( (-1)^{j-1} \mathop \sum \limits_{\begin{subarray}{c}
          {n_1}=1,\dots,{n_j} = 1 \\
          {n_1} \neq \dots \neq {n_j}
        \end{subarray}} ^{N_r}\! \frac{1}{1-\! d_2 \overline{\gamma}_{S_2D}s}\frac{1}{1-\! d_R \overline{\gamma}_{NC,j}s}\Big).
    \end{align}
    The  UPEP $\mathrm{\widetilde{P}_u}\Big(d|\mathbf{W}_d\Big)$ is computed using the MGF method as follows:
    \begin{align} \label{eq:44}
        \mathrm{\widetilde{P}_u}&\Big(d|\mathbf{W}_d\Big) = {\mathop \sum \limits_{j=1}^{N_r}} \Big( (-1)^{j-1} \mathop \sum \limits_{\begin{subarray}{c}
          {n_1}=1,\dots,{n_j} = 1 \\
          {n_1} <\dots< {n_j}
        \end{subarray}} ^{N_r} \Big. \notag \\
    &\Big. \frac{1}{\pi} \int \limits_{0}^{\pi/2} \frac{\sin\theta^4}{\Big(\sin\theta^2 + d_2 \overline{\gamma}_{S_2D}\Big)\Big(\sin\theta^2 + d_R \overline{\gamma}_{{NC}, j}\Big)}d\theta \Big)\notag \\
        =& {\mathop \sum \limits_{j=1}^{N_r}} \Big( (-1)^{j-1} \mathop \sum \limits_{\begin{subarray}{c}
          {n_1}=1,\dots,{n_j} = 1 \\
          {n_1} \neq \dots \neq {n_j}
        \end{subarray}} ^{N_r} \mathcal{I}_1\Big(d_2\overline{\gamma}_{S_2D},d_R\overline{\gamma}_{NC,j}\Big) \Big).
    \end{align}

\item \textbf{Case 3}: $d_2 = 0$. Similar to Case 2 we have $ \mathrm{\widetilde{P}_u}\Big(d|\mathbf{W}_d\Big)$ equals
    \begin{align*} 
       {\mathop \sum \limits_{j=1}^{N_r}} \Big( (-1)^{j-1} \mathop \sum \limits_{\begin{subarray}{c}
          {n_1}=1,\dots,{n_j} = 1 \\
          {n_1} \neq \dots \neq {n_j}
        \end{subarray}} ^{N_r} \mathcal{I}_1\Big(d_1\overline{\gamma}_{S_1D},d_R\overline{\gamma}_{NC,j}\Big) \Big).
    \end{align*}

\item \textbf{Case 4}: $d_1d_2d_R \neq 0$. In this case, there are three weights in the total SNR, resulting in $\gamma_{\Sigma_{NC}} = d_1\gamma_{S_1D} + d_2\gamma_{S_2D} + d_R \gamma_{NC}$. The MGF of $\gamma_{\Sigma_{NC}}$ is given as follows:
    \begin{align}\label{eq:C7}
        \Psi_{\gamma_{\Sigma_{NC}}}(s) &= \Psi_{\gamma_{S_1D}}(d_1s) \times \Psi_{\gamma_{S_2D}}(d_2s) \times \Psi_{\gamma_{{NC}}}(d_Rs)\notag \hfill \\
         &= {\mathop \sum \limits_{j=1}^{N_r}} \Big( (-1)^{j-1} \mathop \sum \limits_{\begin{subarray}{c}
          {n_1}=1,\dots,{n_j} = 1 \\
          {n_1} \neq \dots \neq {n_j}
        \end{subarray}} ^{N_r} \notag \\
       &~~~ \frac{1}{1-d_1 \overline{\gamma}_{S_1D}s} \frac{1}{1- d_2 \overline{\gamma}_{S_2D}s}\frac{1}{1- d_R \overline{\gamma}_{NC,j}s}\Big).
    \end{align}
    Appying the MGF method to compute the UPEP, we have:
    \begin{align} \label{eq:C8}
        &\mathrm{\widetilde{P}_u}\Big(d|\mathbf{W}_d\Big) = {\mathop \sum \limits_{j=1}^{N_r}} \Big( (-1)^{j-1} \mathop \sum \limits_{\begin{subarray}{c}
          {n_1}=1,\dots,{n_j} = 1 \\
          {n_1} \neq \dots \neq {n_j}
        \end{subarray}} ^{N_r} \frac{1}{\pi}\int \limits_{0}^{\pi/2} \notag \\
    &\frac{\sin\theta^6}{(\sin\theta^2\!+\! d_1 \overline{\gamma}_{S_1D}) (\sin\theta^2\!+\! d_2 \overline{\gamma}_{S_2D})(\sin\theta^2\! +\! d_R \overline{\gamma}_{NC,j})}d\theta \Big)\notag \\
        &= {\mathop \sum \limits_{j=1}^{N_r}} \Big( (-1)^{j-1}\! \mathop \sum \limits_{\begin{subarray}{c}
          {n_1}=1,\dots,{n_j} = 1 \\
          {n_1} \neq \dots \neq {n_j}
        \end{subarray}}^{N_r}\! \mathcal{I}_2(d_1\overline{\gamma}_{S_1D}, d_2\overline{\gamma}_{S_2D},d_R\overline{\gamma}_{NC,j}) \Big),
    \end{align}
    where $\mathcal{I}_2(a,b,c)$ has been defined in Theorem~\ref{theorem:3}.

\end{itemize}
Combining four cases above gives Theorem~\ref{theorem:3}.
\section{Proof of Theorem~\ref{theorem:4}}
\label{App:D}
\renewcommand{\theequation}{D.\arabic{equation}}
\setcounter{equation}{0}
Similar to Appendix~\ref{App:B}, we employ the upper bound of UPEP to derive diversity order for NCC:
\begin{align}\label{eq:D2}
\tau_n \geq - \lim_{\overline{\gamma} \rightarrow \infty} \Big(\frac{\log \Psi_{\gamma_{\Sigma_{NC}}}(1/2)}{\log \overline{\gamma}}\Big).
\end{align}
We consider four cases:

\begin{itemize}
\item \textbf{Case 1}: $d_R = 0$. In this case, all weights locate in the source-destination channels, resulting in $\Psi_{\gamma_{\Sigma_{NC}}}\Big(1/2\Big) = \Psi_{\gamma_{S_1D}}\Big(d_1/2\Big)\times \Psi_{\gamma_{S_2D}}\Big(d_2/2\Big)$. The diversity order in this case is given by:
    \begin{align}\label{eq:D3}
        \tau_n &\geq -\lim_{\overline{\gamma} \rightarrow \infty} \frac{\log \Psi_{\gamma_{S_1D}}\Big(d_1/2\Big)}{\log \overline{\gamma}} -\lim_{\overline{\gamma} \rightarrow \infty} \frac{\log \Psi_{\gamma_{S_2D}}\Big(d_2/2\Big)}{\log \overline{\gamma}} \notag \\
        & \geq -\lim_{\overline{\gamma} \rightarrow \infty}\Big(\frac{\Big(1 + d_1\overline{\gamma}_{S_1D}/2\Big)^{-1}} {\log \overline{\gamma}}\Big) \notag \\
        &~~~ - \lim_{\overline{\gamma} \rightarrow \infty}\Big(\frac{\Big(1 + d_2\overline{\gamma}_{S_2D}/2\Big)^{-1}} {\log \overline{\gamma}}\Big) \notag \\
        &= 1 + 1 = 2. 
    \end{align}
    This is enough to say the UPEP has diversity order of 2 when $d_R = 0$ and we can write $\mathrm{\widetilde{P}_u}\Big(d|d_R = 0\Big) \circeq \mathrm{SNR}^{-2}$.
\item \textbf{Case 2}: $d_1 = 0$. The MGF of the total SNR in this case has a form of $\Psi_{\gamma_{\Sigma_{NC}}}\Big(1/2\Big) = \Psi_{\gamma_{S_2D}}\Big(d_2/2\Big)\times \Psi_{\gamma_{NC}}\Big(d_R/2\Big)$. Consequently, the diversity order is given as follows:
    \begin{align}\label{eq:D4}
        \tau_n &\geq -\lim_{\overline{\gamma} \rightarrow \infty} \frac{\log \Psi_{\gamma_{S_2D}}\Big(d_2/2\Big)}{\log \overline{\gamma}} \underbrace {-\lim_{\overline{\gamma} \rightarrow \infty} \frac{\log \Psi_{\gamma_{NC}}\Big(d_R/2\Big)}{\log \overline{\gamma}}}_{\mathcal{J}} \notag \\
        &= 1 +  \mathcal{J},
    \end{align}
    where $\mathcal{J}$ is the diversity order of the best relayed signal without the direct link, which equals diversity order of the best relay selection for two-way relay channels. It has shown in \cite{Yonghui2010} that this diversity order is equal to $N_r$. Therefore we have the system diversity order when $d_1 = 0$ is equal to $N_r + 1$. In order words, $\mathrm{\widetilde{P}_u}\Big(d|d_1 = 0\Big) \circeq \mathrm{SNR}^{-N_r-1}$.
\item \textbf{Case 3}: $d_2 = 0$. Similar to case 2 we have the diversity order equals $N_r+1$.
\item \textbf{Case 4}: $d_1d_2d_R \neq 0$. In this case the MGF of $\gamma_{\Sigma_{NC}}$ is a product of three terms:
    \begin{align}\label{eq:D5}
    &\Psi_{\gamma_{\Sigma_{NC}}}\Big(1/2\Big) \\
    &~~~= \Psi_{\gamma_{S_1D}}\Big(d_1/2\Big)\times \Psi_{\gamma_{S_2D}}\Big(d_2/2\Big)\times \Psi_{\gamma_{NC}}\Big(d_R/2\Big). \notag
    \end{align}
    Substituting \eqref{eq:D5} into \eqref{eq:D2} we have
    \begin{align}\label{eq:D6}
        \tau_n &\geq -\lim_{\overline{\gamma} \rightarrow \infty} \frac{\log \Psi_{\gamma_{S_1D}}\Big(d_1/2\Big)}{\log \overline{\gamma}} -\lim_{\overline{\gamma} \rightarrow \infty} \frac{\log \Psi_{\gamma_{S_2D}}\Big(d_2/2\Big)}{\log \overline{\gamma}}\notag \\
        &~~~~ \underbrace{-\lim_{\overline{\gamma} \rightarrow \infty} \frac{\log \Psi_{\gamma_{NC}}\Big(d_R/2\Big)}{\log \overline{\gamma}}}_{\mathcal{J}} \notag \\
        &= 1 + 1 + \mathcal{J} = N_r + 2.
    \end{align}
    We can write $\mathrm{\widetilde{P}_u}\Big(d|d_1d_2d_R \neq 0\Big) \circeq \mathrm{SNR}^{-N_r-2}$.
\end{itemize}
From four cases above we have Theorem~\ref{theorem:4} proved.

\end{document}